\documentclass{webofc}
\usepackage[varg]{txfonts}   % Web of Conferences font
\usepackage{hyperref}
\usepackage{url}
\usepackage{amsmath,bm}
\usepackage{tikz}
\usetikzlibrary{arrows.meta,decorations.pathmorphing}
\hypersetup{colorlinks=true,citecolor=blue,urlcolor=blue,linkcolor=blue}
\DeclareMathOperator{\re}{Re}
\DeclareMathOperator{\im}{Im}

\newcommand\bsub{\begin{subequations}}
\newcommand\esub{\end{subequations}}
\newcommand{\rr}[3]{r^{#1}_{\substack{#2\\ #3}}}
\newcommand{\rt}[3]{\tilde r^{#1}_{\substack{#2\\ #3}}}
\newcommand{\rhoo}[3]{\rho^{#1}_{\substack{#2\\ #3}}}

\begin{document}
%
% Alternative titles under consideration:
%  - Meson photoproduction: from the $a_2(1320)$ to $\rho^-\Delta^{++}$ spin correlations
%  - Meson photoproduction at Jefferson Lab: from two-meson final states to meson-baryon spin-density matrices
\title{Meson photoproduction at Jefferson Lab: from two-meson final states to meson-baryon spin-density matrices}
\author{\firstname{Vincent} \lastname{Mathieu}\inst{1}\fnsep\thanks{\email{vmathieu@ub.edu}}}

\institute{Departament de F\'isica Qu\`antica i Astrof\'isica and Institut de Ci\`encies del Cosmos (ICCUB), \\ Universitat de Barcelona, E-08028 Barcelona, Spain}

\abstract{Recent progress in the understanding of meson photoproduction, obtained from the joint effort of the Joint Physics Analysis Center (JPAC) and the GlueX and CLAS experiments at Jefferson Lab, is reviewed. The photoproduction of the $\eta\pi$ system is now well understood, from the $a_2(1320)$ resonance region to the high-mass region dominated by double Regge exchanges. Since the lightest hybrid meson, the $\pi_1(1600)$, is expected to decay predominantly into $b_1\pi$, reactions with a recoiling $\Delta^{++}$ are promising channels for exotic searches. As a first step towards $\gamma p \to b_1 \pi \Delta^{++}$, the complete angular distribution of $\gamma p \to \rho^-\Delta^{++}$ with a linearly polarized photon beam is derived. It involves 56 double-index spin-density matrix elements, which describe the correlations between the $\rho$ and the $\Delta$ decays. When the production amplitudes factorize, as is the case for the exchange of a single Regge pole, the intensity reduces to the product of the standard $\rho$ and $\Delta$ decay distributions and depends on only 14 parameters.}
\maketitle
%
%=========================================================================
\section{Introduction}\label{sec:intro}
%=========================================================================
Quantum chromodynamics allows for hadrons beyond the conventional quark-antiquark mesons and three-quark baryons. Among them, hybrid mesons, in which the gluon field is excited and contributes to the quantum numbers, are prime candidates to study the role of gluons in the hadron spectrum. Lattice QCD predicts that the lightest isovector hybrid is the $\pi_1$, with the exotic quantum numbers $J^{PC}=1^{-+}$~\cite{Dudek:2010wm}. Its couplings to $\eta\pi$, $\eta'\pi$, $\rho\pi$, $b_1\pi$, \ldots\ have also been computed on the lattice, and the $b_1\pi$ channel is expected to dominate its width~\cite{Woss:2020ayi}. The $\eta^{(\prime)}\pi$ final states are experimentally the cleanest, since any odd partial wave in these systems carries exotic quantum numbers.

The main goal of the GlueX experiment, which uses the linearly polarized photon beam of Jefferson Lab, is the search for hybrid mesons. This requires a detailed understanding of the production mechanisms. In this contribution, I summarize the joint findings of JPAC and of the GlueX and CLAS collaborations on $\eta\pi$ photoproduction (Sec.~\ref{sec:etapi}). I then review the spin-density matrix elements (SDME) of the $\rho$ and the $\Delta$ (Sec.~\ref{sec:SDME}), and derive the complete intensity of $\gamma p\to\rho^-\Delta^{++}$ in terms of double-index SDME (Sec.~\ref{sec:rhoDelta}), as well as its factorized form (Sec.~\ref{sec:factorization}). These expressions are used in the GlueX analysis of $\gamma p\to\rho^-\Delta^{++}$ presented in Refs.~\cite{Afzal:MESON2026,Herrmann:MESON2026}.

%=========================================================================
\section{Photoproduction of $\eta\pi$}\label{sec:etapi}
%=========================================================================
\paragraph{Tensor mesons.}
The $\eta\pi^0$ mass spectrum measured by GlueX is dominated by the $a_0(980)$ and the $a_2(1320)$. The latter is an ideal benchmark for the production mechanism. In Ref.~\cite{Mathieu:2020zpm}, we developed a model for tensor meson photoproduction based on the exchange of vector ($\rho$ and $\omega$) and axial-vector mesons, with couplings constrained by the radiative decays. In this model, the $a_2(1320)$ is produced mostly by $\omega$ exchange, with a smaller $\rho$ contribution. The situation is reversed for the $f_2(1270)$. The model compares well with the CLAS measurements of $a_2(1320)$~\cite{CLAS:2020rdz} and $f_2(1270)$~\cite{CLAS:2020ngl} photoproduction. The GlueX collaboration then measured the polarized cross section of $a_2(1320)$ photoproduction at $E_\gamma\sim 8.5$~GeV and separated the two reflectivity components~\cite{GlueX:2025kma}. At high energies, the positive (negative) reflectivity corresponds to natural (unnatural) exchanges, \textit{i.e.}, to vector (axial-vector) mesons. The positive reflectivity dominates the cross section at small momentum transfer, in agreement with the JPAC prediction. All $D$-wave components have been extracted~\cite{GlueX:2025kma}.

\paragraph{High-mass region.}
Above the resonance region, the $\eta\pi$ events accumulate at forward and backward angles in the $\eta\pi$ rest frame. This behavior is characteristic of double Regge exchanges, in which either the $\pi^0$ or the $\eta$ is emitted at the photon vertex. In Ref.~\cite{Montana:2025asi}, we built a double Regge model including the leading $\rho$ and $\omega$ exchanges. All couplings are fixed from the radiative decays $\rho,\omega\to\gamma\pi^0,\gamma\eta$ and $\omega\to\rho\pi$, or from SU(3) relations, so the model has no free parameters. It reproduces the CLAS $\eta\pi$ mass spectrum above $m_{\eta\pi}\sim1.6$~GeV. The model predicts a forward-backward asymmetry, which can only arise from the interference between even and odd partial waves. Odd, and thus exotic, waves are therefore naturally present in the non-resonant $\eta^{(\prime)}\pi$ spectrum. The asymmetry is much larger in $\eta'\pi$ than in $\eta\pi$~\cite{Montana:2025asi}.

\paragraph{Moments and upper limits.}
The formalism relating the angular moments of the $\eta\pi$ system to the partial waves and to the beam polarization was established in Ref.~\cite{Mathieu:2019fts} for a linearly polarized beam, and recently extended to elliptical polarization~\cite{Glazier:2025emr}. The GlueX moments of the $\eta\pi^0$ system have been extracted and are being finalized (GlueX collaboration, in preparation). Finally, GlueX has set an upper limit on the photoproduction cross section of the $\pi_1(1600)$ in the $\eta^{(\prime)}\pi$ channels~\cite{GlueX:2024erj}. Since the $\pi_1(1600)$ decays dominantly into $b_1\pi$~\cite{Woss:2020ayi}, the reaction $\gamma p\to\pi_1^-\Delta^{++}\to (b_1^-\pi^0)\Delta^{++}$, with $b_1^-\to\omega\pi^-$ and $\Delta^{++}\to p\pi^+$, is a promising channel. A first step is the simplest meson-baryon channel, $\gamma p\to\rho^-\Delta^{++}$, which is the subject of the rest of this contribution.

%=========================================================================
\section{Single spin-density matrices}\label{sec:SDME}
%=========================================================================
When a single resonance of known spin is produced, its decay angular distribution is fully determined by its SDME. For a linearly polarized photon beam, with polarization $P_\gamma$ and angle $\Phi$ between the polarization vector and the production plane, the intensity reads
\begin{align}
    I(\Omega,\Phi) & = I^0(\Omega) - P_\gamma I^1(\Omega)\cos2\Phi - P_\gamma I^2(\Omega)\sin 2\Phi .
    \label{eq:Ipol}
\end{align}
For a vector meson decaying into two pions, $\rho\to\pi\pi$, one has $I^\alpha(\Omega) = \tfrac{3}{4\pi}W^\alpha_\rho(\Omega)$ with~\cite{Schilling:1969um}
\bsub\label{eq:Wrho}\begin{align}
    W^{0,1}_\rho(\Omega) & = r^{\rho,0,1}_{00} \cos^2\theta + r^{\rho,0,1}_{11} \sin^2\theta - r^{\rho,0,1}_{1-1} \cos 2\phi \sin^2\theta - \sqrt{2}\, r^{\rho,0,1}_{10} \cos\phi \sin2 \theta ,
    \\
    W^{2}_\rho(\Omega) & = r^{\rho,2}_{1-1} \sin 2\phi \sin^2\theta + \sqrt{2}\, r^{\rho,2}_{10} \sin\phi \sin2 \theta ,
\end{align}\esub
where $r^{\rho,\alpha}_{mm'} = \re\rho^\alpha_{mm'}$ for $\alpha=0,1$ and $r^{\rho,2}_{mm'} = \im\rho^2_{mm'}$. The nine SDME of the $\rho^0$ have been measured by GlueX at $E_\gamma=8.2$--$8.8$~GeV~\cite{GlueX:2023fcq}. They confirm the dominance of natural exchanges predicted by JPAC~\cite{Mathieu:2018xyc}.

Similarly, the decay $\Delta^{++}\to p\pi^+$ is described by
\begin{align}\nonumber
    W^{0,1}_\Delta(\Omega) & = r^{\Delta,0,1}_{11} \left( \frac{1}{3} +\cos^2\theta\right) + r^{\Delta,0,1}_{33} \sin^2 \theta - \frac{2}{\sqrt{3}} \left[ r^{\Delta,0,1}_{31} \cos\phi \sin2\theta + r^{\Delta,0,1}_{3-1} \cos2\phi \sin^2\theta \right] ,
    \\
    W^{2}_\Delta(\Omega) & = \frac{2}{\sqrt{3}} \left[ r^{\Delta,2}_{31} \sin\phi \sin2\theta + r^{\Delta,2}_{3-1} \sin2\phi \sin^2\theta \right] ,
    \label{eq:WDelta}
\end{align}
where the $\Delta$ helicities are denoted by twice their value, \textit{e.g.}, $r^{\Delta,0}_{31}\equiv \re\rho^{0}_{\frac32\frac12}$. GlueX has measured the nine SDME of the $\Delta^{++}$ in $\gamma p\to\pi^-\Delta^{++}$~\cite{GlueX:2024dbr}. The early JPAC model~\cite{JointPhysicsAnalysisCenter:2017del} was recently extended to a global description of the cross section, the beam asymmetry and all SDME~\cite{Shastry:2026vlf}. This provided the first determination of the vector and tensor couplings of the $N\Delta$ transition~\cite{Shastry:2026mkk}.

%=========================================================================
\section{The reaction $\gamma p\to\rho^-\Delta^{++}$}\label{sec:rhoDelta}
%=========================================================================
\begin{figure}[t]
\centering
\sidecaption
\begin{tikzpicture}[scale=0.75,>={Stealth[length=2.2mm]},thick]
  \coordinate (T) at (0,0.9);  \coordinate (B) at (0,-0.9);
  \coordinate (R) at (1.8,0.9); \coordinate (D) at (1.8,-0.9);
  \draw[decorate,decoration={snake,amplitude=2pt,segment length=6pt}] (-1.8,0.9) -- (T);
  \node[left] at (-1.8,0.9) {$\gamma$};
  \draw[->] (-1.8,-0.9) -- (-0.8,-0.9); \draw (-0.8,-0.9) -- (B);
  \node[left] at (-1.8,-0.9) {$p$};
  \draw[blue!70!black,line width=2.2pt,decorate,decoration={snake,amplitude=2.5pt,segment length=9pt}] (T) -- (B);
  \node[blue!70!black] at (-0.45,0) {$\mathcal{R}$};
  \draw[blue!70!black,double,double distance=1.6pt,line width=0.9pt] (T) -- (R);
  \draw[blue!70!black,double,double distance=1.6pt,line width=0.9pt] (B) -- (D);
  \node[blue!70!black,above] at (0.9,0.95) {$\rho^-$};
  \node[blue!70!black,below] at (0.9,-0.95) {$\Delta^{++}$};
  \foreach \p in {T,B,R,D} \fill[blue!70!black] (\p) circle (4.5pt);
  \draw[->] (R) -- ++(1.1,0.45) node[right] {$\pi^0$};
  \draw[->] (R) -- ++(1.1,-0.45) node[right] {$\pi^-$};
  \draw[->] (D) -- ++(1.1,0.45) node[right] {$\pi^+$};
  \draw[->] (D) -- ++(1.1,-0.45) node[right] {$p$};
\end{tikzpicture}
\caption{The reaction $\gamma p\to\rho^-\Delta^{++}\to(\pi^0\pi^-)(\pi^+p)$. At high energies, the production is dominated by the exchange of Regge poles $\mathcal{R}$ in the $t$-channel.}
\label{fig:rhoDelta}
\end{figure}
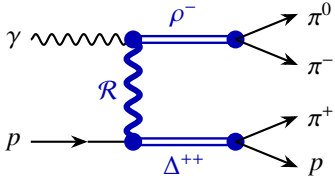

\subsection{Amplitudes}
The final state $\pi^0\pi^-\pi^+p$ of the reaction in Fig.~\ref{fig:rhoDelta} is described by $3\times4-4=8$ variables. We choose the Mandelstam variables $s$ and $t$ of the quasi-two-body reaction $\gamma p\to\rho^-\Delta^{++}$, the invariant masses $m_{\pi\pi}$ and $m_{p\pi}$, and two pairs of angles: $\Omega_\pi$, the angles of the $\pi^-$ in the $\pi\pi$ rest frame, and $\Omega_p$, the angles of the proton in the $p\pi$ rest frame. We use Gottfried--Jackson frames: the $z$ axis is along the beam in the $\pi\pi$ rest frame and along the target in the $p\pi$ rest frame. In both frames the $y$ axis is normal to the production plane. Restricting the $\pi\pi$ system to the $P$-wave and the $p\pi$ system to the $\frac32^+$ wave, the amplitude reads
\begin{align}
    A_{\lambda_\gamma,\lambda_1,\lambda_2} & = \frac{1}{\sqrt{\pi}}\sum_{m=-1}^{1}\sum^{3/2}_{\lambda = -3/2} V_{\lambda_\gamma,  m;\lambda_1,\lambda}(s,t)\, F(m_{\pi\pi})\, Y^{1}_{m}(\Omega_\pi)\,
   \tilde F(m_{p\pi})\, D^{3/2*}_{\lambda, \lambda_2}(\Omega_{p}) .
   \label{eq:amp}
\end{align}
Here $\lambda_\gamma$, $\lambda_1$ and $\lambda_2$ are the helicities of the photon, the target and the recoil proton, and $m$ ($\lambda$) is the spin projection of the $\rho$ ($\Delta$). The functions $F$ and $\tilde F$ are the lineshapes of the $\rho$ and the $\Delta$. Parity conservation implies $V_{-1,-m;-\lambda_1,-\lambda} = (-1)^{m}(-1)^{\lambda-\lambda_1}V_{1,m;\lambda_1,\lambda}$. The production is therefore described by 24 independent complex functions of $s$ and $t$.

\subsection{Double-index spin-density matrix elements}
The intensity takes the form of Eq.~\eqref{eq:Ipol}, with $\Omega\to(\Omega_\pi,\Omega_p)$ and
\begin{align}
    I^\alpha(\Omega_\pi,\Omega_p) & = \frac{2\kappa}{\pi}\left|F \tilde F\right|^2
    \sum_{m,m'}\sum_{\lambda,\lambda'} Y^1_m(\Omega_\pi)\, \rhoo{\alpha}{mm'}{\lambda\lambda'}(s,t)\, Y^{1*}_{m'}(\Omega_\pi)
    \sum_{\lambda_2} D^{3/2*}_{\lambda,\lambda_2}(\Omega_p) D^{3/2}_{\lambda',\lambda_2}(\Omega_p) ,
\end{align}
where $\kappa$ contains the phase space. The double-index SDME are defined by
\begin{align}
    \rhoo{\alpha}{mm'}{\lambda\lambda'}(s,t) & = \frac{1}{2}\sum_{\lambda_\gamma,\lambda'_\gamma}\sum_{\lambda_1}
    V_{\lambda_\gamma, m;\lambda_1,\lambda}(s,t)\, \sigma^\alpha_{\lambda_\gamma,\lambda'_\gamma}\, V^{*}_{\lambda'_\gamma, m';\lambda_1,\lambda'}(s,t) ,
\end{align}
with $\sigma^0$ the identity and $\sigma^{1,2,3}$ the Pauli matrices. They are hermitian, $\rhoo{\alpha}{m'm}{\lambda'\lambda} = (\rhoo{\alpha}{mm'}{\lambda\lambda'})^*$. Parity implies $\rhoo{\alpha}{-m\,-m'}{-\lambda\,-\lambda'} = \pm(-1)^{m-m'}(-1)^{\lambda-\lambda'}\rhoo{\alpha}{mm'}{\lambda\lambda'}$, with the $+$ ($-$) sign for $\alpha=0,1$ ($\alpha=2,3$). Pulling out the factor $(3/2)\times(3/4\pi)$, we write $I^\alpha = 9\kappa/(4\pi^2)|F\tilde F|^2\, W^\alpha(\Omega_\pi,\Omega_p)$. We expand the double angular distributions in the $\rho$ decay angles $\Omega_\pi=(\theta_\pi,\phi_\pi)$:
\bsub\label{eq:Wfull}\begin{align}\nonumber
    W^{0,1}(\Omega_\pi,\Omega_{p}) & =
     \cos^2\theta_\pi  \left[ W_{00}^{0,1} - \overline W_{00}^{0,1} \right] + \sin^2\theta_\pi \left[ W_{11}^{0,1} - \overline W_{11}^{0,1} \right]
    \\ \nonumber
    & -\sin^2\theta_\pi \left(\cos 2\phi_\pi \left[ W_{1-1}^{0,1} - \overline W_{1-1}^{0,1} \right]
    + \sin 2\phi_\pi \, \widetilde W_{1-1}^{0,1} \right)
    \\
    &
    - \sqrt{2} \sin2\theta_\pi \left(
    \cos\phi_\pi \left[ W_{10}^{0,1} - \overline W_{10}^{0,1} \right]  + \sin\phi_\pi \, \widetilde W_{10}^{0,1}  \right) ,
    \label{eq:W01}
    \\ \nonumber
    W^2(\Omega_\pi,\Omega_{p})& =
     \cos^2\theta_\pi \, \widetilde W_{00}^2 + \sin^2\theta_\pi \, \widetilde W_{11}^2
     +\sin^2\theta_\pi \left(\sin 2\phi_\pi \left[ W_{1-1}^{2} - \overline W_{1-1}^{2} \right]
    - \cos 2\phi_\pi \, \widetilde W_{1-1}^{2} \right)
    \\
    &
    + \sqrt{2} \sin2\theta_\pi \left(\sin\phi_\pi \left[ W_{10}^{2} - \overline W_{10}^{2} \right]  + \cos\phi_\pi \, \widetilde W_{10}^{2}
    \right) .
    \label{eq:W2}
\end{align}\esub
The functions $W^\alpha_{mm'}$, $\overline W^\alpha_{mm'}$ and $\widetilde W^\alpha_{mm'}$ depend on the $\Delta$ decay angles $\Omega_p=(\theta,\phi)$ through the three structures of the $\Delta$ decay distribution:
\bsub\label{eq:Waux}\begin{align}
    W_{mm'}^{\alpha}(\Omega) & = \rr{\alpha}{mm'}{3 3} \sin^2 \theta+ \rr{\alpha}{mm'}{1 1} \left( \frac{1}{3} +\cos^2\theta\right) ,
    \\
    \overline W_{mm'}^{\alpha}(\Omega) & =
    \frac{2}{\sqrt{3}} \left[ \rr{\alpha}{mm'}{3 1} \cos\phi \sin2\theta +  \rr{\alpha}{mm'}{3 -1} \cos2\phi \sin^2\theta \right] ,
    \\
    \widetilde W_{mm'}^{\alpha}(\Omega) & = \frac{2}{\sqrt{3}} \left[ \rt{\alpha}{mm'}{3 1} \sin\phi \sin2\theta + \rt{\alpha}{mm'}{3 -1} \sin2\phi \sin^2\theta \right] .
\end{align}\esub
For $(m,m')=(0,0)$, $(1,1)$ and $(1,-1)$, the real coefficients $r$ and $\tilde r$ are related to the double-index SDME by
\bsub\label{eq:rdef}\begin{align}
    \rr{0,1}{mm'}{\lambda \lambda'} & = \frac{1}{2}\re \left(\rhoo{0,1}{mm'}{\lambda \lambda'}+(-1)^{\lambda-\lambda'}\rhoo{0,1}{mm'}{-\lambda\, -\lambda'}\right) ,
    &
    \rt{0,1}{mm'}{\lambda \lambda'} & =  \frac{1}{2}\re \left(\rhoo{0,1}{mm'}{\lambda \lambda'}-(-1)^{\lambda-\lambda'}\rhoo{0,1}{mm'}{-\lambda\, -\lambda'}\right) ,
    \\
    \rr{2}{mm'}{\lambda \lambda'} & = \frac{1}{2}\im \left(\rhoo{2}{mm'}{\lambda \lambda'}+(-1)^{\lambda-\lambda'}\rhoo{2}{mm'}{-\lambda\, -\lambda'}\right) ,
    &
    \rt{2}{mm'}{\lambda \lambda'} & =  \frac{1}{2}\im \left(\rhoo{2}{mm'}{\lambda \lambda'}-(-1)^{\lambda-\lambda'}\rhoo{2}{mm'}{-\lambda\, -\lambda'}\right) ,
\end{align}\esub
while for $(m,m')=(1,0)$ they read, for $\alpha=0,1$,
\bsub\label{eq:rdef10}\begin{align}
    \rr{\alpha}{10}{\lambda\lambda} & = \frac{1}{2}\re \left(\rhoo{\alpha}{10}{\lambda\lambda}+\rhoo{\alpha}{10}{-\lambda\, -\lambda}\right) , \qquad \lambda=\tfrac32,\tfrac12,
    \\
    \overset{(\sim)}{r}{}^{\alpha}_{\substack{10\\ 3 1}} & = \frac{1}{4}\re \left[\left( \rhoo{\alpha}{10}{3 1}  - \rhoo{\alpha}{10}{-1 -3} \right) \pm \left(\rhoo{\alpha}{10}{1 3}  - \rhoo{\alpha}{10}{-3 -1}\right)  \right] ,
    \\
    \overset{(\sim)}{r}{}^{\alpha}_{\substack{10\\ 3 -1}} & = \frac{1}{4}\re  \left[ \left(\rhoo{\alpha}{10}{3 -1}  + \rhoo{\alpha}{10}{1 -3} \right) \pm \left( \rhoo{\alpha}{10}{-1 3}  + \rhoo{\alpha}{10}{-3 1} \right)   \right] ,
\end{align}\esub
where $r$ ($\tilde r$) takes the upper (lower) sign. The $\alpha=2$ coefficients follow from Eq.~\eqref{eq:rdef10} with $\re\to\im$ and an additional overall minus sign for $\tilde r^2$.

Parity and hermiticity leave 20 independent coefficients for each of $\alpha=0,1$: four $r$ for each of $(m,m')=(0,0)$ and $(1,1)$, whose $\tilde r$ vanish, and four $r$ and two $\tilde r$ for each of $(1,-1)$ and $(1,0)$. For $\alpha=2$, only the $\tilde r$ contribute for $(0,0)$ and $(1,1)$, giving 16 coefficients in total. The complete intensity thus depends on $20+20+16=56$ real coefficients at each $(s,t)$. After integration over both decay angles, only $\rr{0}{00}{33}+\rr{0}{00}{11}+2(\rr{0}{11}{33}+\rr{0}{11}{11})$ survives. It is proportional to the differential cross section.

The expressions~\eqref{eq:Wfull} and~\eqref{eq:Waux} make the connection with the single SDME of Sec.~\ref{sec:SDME} explicit. Integrating over $\Omega_p$ removes $\overline W$ and $\widetilde W$ and gives $W^\alpha_\rho$ with $r^{\rho,\alpha}_{mm'}\propto\rr{\alpha}{mm'}{33}+\rr{\alpha}{mm'}{11}$. Integrating over $\Omega_\pi$ gives $W^\alpha_\Delta$ with $r^{\Delta,\alpha}_{\lambda\lambda'}\propto\rr{\alpha}{00}{\lambda\lambda'}+2\rr{\alpha}{11}{\lambda\lambda'}$ for $\alpha=0,1$. The genuinely new information lies in the correlations between the two decays.

\subsection{Reflectivity basis}
Following Ref.~\cite{Mathieu:2019fts}, we introduce the reflectivity amplitudes
$^{(\epsilon)}V_{m;\lambda_1,\lambda} = \frac{1}{2}\left[V_{1,m; \lambda_1, \lambda} - \epsilon(-1)^m V_{-1,-m; \lambda_1, \lambda} \right]$. At high energies, the reflectivity $\epsilon=\pm$ coincides with the naturality of the exchanged Regge pole. Parity relates the amplitudes with $\lambda_1=-\frac12$ to those with $\lambda_1=+\frac12$. This leaves 12 positive-reflectivity and 12 negative-reflectivity amplitudes. The intensity can be written as a manifestly positive sum of squared moduli of these 24 amplitudes. %This is how it is implemented in the fits of Refs.~\cite{Afzal:MESON2026,Herrmann:MESON2026}. 
In terms of reflectivity amplitudes, the combinations $\rr{0}{mm'}{\lambda\lambda'}\pm\rr{1}{m\,-m'}{\lambda\lambda'}$ are bilinear in amplitudes of a single reflectivity, either positive or negative depending on the sign. The coefficients $\tilde r$, on the other hand, are pure interferences between the two reflectivities.

This differs from the photoproduction of two pseudoscalar mesons. When the recoiling baryon is a nucleon, or when the $\Delta$ decay angles are integrated over, the two reflectivities contribute incoherently to the $\alpha=0,1$ intensities~\cite{Mathieu:2019fts}. With the $\Delta$ decay distribution measured, the angular functions $\sum_{\lambda_2}D^{3/2*}_{\lambda,\lambda_2}D^{3/2}_{\lambda',\lambda_2}$ are complex for $|\lambda|\neq|\lambda'|$. Interferences between reflectivities then survive through the $\widetilde W$ terms, which are odd in $\phi$.

%=========================================================================
\section{Factorization}\label{sec:factorization}
%=========================================================================
Preliminary results of a Regge model indicate that $\gamma p\to\rho^-\Delta^{++}$ is dominated by the exchange of the natural-parity $\rho$ trajectory (JPAC collaboration, in preparation). For a single Regge pole, the residues factorize into a meson vertex and a baryon vertex, and the amplitudes take the form
\begin{align}
    V_{\lambda_\gamma, m; \lambda_1,\lambda}(s,t) & = \beta_{\lambda_\gamma m}(t)\; \gamma_{\lambda_1\lambda}(t)\; f(s,t) .
\end{align}
For an exchange of naturality $\eta=P(-1)^J$, each vertex satisfies its own parity relation,
\begin{align}
    \beta_{-\lambda_\gamma, -m}(t) & = \eta(-1)^{\lambda_\gamma-m}\beta_{\lambda_\gamma m}(t) ,
    &
    \gamma_{-\lambda_1, -\lambda}(t) & = \eta(-1)^{\lambda_1+\lambda}\gamma_{\lambda_1\lambda}(t) .
\end{align}
The double-index SDME then factorize into a meson and a baryon SDME, $\rhoo{\alpha}{mm'}{\lambda\lambda'}=\rho^{\rho,\alpha}_{mm'}\,\rho^{\Delta}_{\lambda\lambda'}$, where the baryon factor does not depend on the photon polarization. As a consequence, all the coefficients $\tilde r$ vanish and $\rr{\alpha}{mm'}{\lambda\lambda'}=r^{\rho,\alpha}_{mm'}\,r^{\Delta}_{\lambda\lambda'}$. The intensity reduces to the product of the standard decay distributions~\eqref{eq:Wrho} and~\eqref{eq:WDelta},
\begin{align}
    I(\Phi,\Omega_\pi,\Omega_p) & = \frac{9\kappa}{4\pi^2}\left|F\tilde F\right|^2
    \left[ W^0_\rho(\Omega_\pi) - P_\gamma W^1_\rho(\Omega_\pi)\cos2\Phi - P_\gamma W^2_\rho(\Omega_\pi)\sin2\Phi \right] W_\Delta(\Omega_p) ,
    \label{eq:Ifact}
\end{align}
where $W_\Delta\equiv W^0_\Delta$ is given by Eq.~\eqref{eq:WDelta} with the coefficients $r^\Delta_{\lambda\lambda'}$. The factorized intensity depends on $4+4+2$ meson coefficients and 4 baryon coefficients, \textit{i.e.}, 14 parameters instead of 56 (one of them can be absorbed in the overall normalization). In terms of the vertex functions, with $\beta_m\equiv\beta_{1m}$ and $\gamma_{2\lambda}\equiv\gamma_{\frac12\lambda}$ taken real, we find
\bsub\label{eq:rfact}\begin{align}
    r^{\rho,0}_{00} & = \beta_0^2 ,
    &
    r^{\rho,0}_{11} & = \tfrac{1}{2}\left(\beta_1^2 + \beta_{-1}^2\right) ,
    &
    r^{\rho,0}_{10} & = \tfrac{1}{2}\beta_0(\beta_1-\beta_{-1}) ,
    \\
    r^{\rho,0}_{1-1} & = \beta_1\beta_{-1} ,
    &
    r^{\rho,2}_{10} & = \tfrac{\eta}{2} \beta_0 \left(\beta_1+\beta_{-1}\right) ,
    &
    r^{\rho,2}_{1-1} & = -\tfrac{\eta}{2}  \left(\beta^2_1-\beta^2_{-1}\right) ,
    \\
    r^\Delta_{33} & = \gamma^2_{3} + \gamma^2_{-3} ,
    &
    r^\Delta_{31} & = \gamma_{3}\gamma_1 - \gamma_{-3}\gamma_{-1} ,
    \\
    r^\Delta_{11} & = \gamma^2_{1} + \gamma^2_{-1} ,
    &
    r^\Delta_{3-1} & = \gamma_{3}\gamma_{-1} + \gamma_{-3}\gamma_{1} ,
\end{align}\esub
and $r^{\rho,1}_{00} = -\eta\, r^{\rho,0}_{00}$, $r^{\rho,1}_{10} = -\eta\, r^{\rho,0}_{10}$, $r^{\rho,1}_{11} = \eta\, r^{\rho,0}_{1-1}$, $r^{\rho,1}_{1-1} = \eta\, r^{\rho,0}_{11}$. These last relations are the usual signatures of an exchange of definite naturality. The fit with the factorized intensity~\eqref{eq:Ifact} does not impose them. Comparing the full fit, with 56 coefficients, and the factorized one, with 14, provides a direct test of the factorization of the production mechanism. This comparison, performed on GlueX data, is presented in Refs.~\cite{Afzal:MESON2026,Herrmann:MESON2026}. Similar observations were reported for $\gamma p\to K^*\Lambda$~\cite{Li:MESON2026}.

%=========================================================================
\section{Summary and outlook}\label{sec:summary}
%=========================================================================
JPAC and the GlueX and CLAS collaborations are building a comprehensive understanding of meson photoproduction, channel by channel. The photoproduction of $\eta\pi$ is well understood, from the $a_2(1320)$ to the high-mass region~\cite{Mathieu:2020zpm,GlueX:2025kma,Montana:2025asi}. In view of the dominant $\pi_1\to b_1\pi$ decay, I derived the complete intensity of $\gamma p\to\rho^-\Delta^{++}$, the simplest meson-baryon channel, in terms of 56 double-index SDME, and its factorized form with 14 parameters. The next steps are $\gamma p\to b_1^-\Delta^{++}$, to identify the $b_1$, and ultimately $\gamma p\to b_1^-\pi^0\Delta^{++}$ for the exotic $\pi_1$, as well as the extension to $K^*\Lambda$.

\begin{acknowledgement}
I thank my colleagues from JPAC and from the GlueX and CLAS collaborations for fruitful collaboration, in particular F.~Afzal, N.~Herrmann and V.~Shastry. I am a Serra H\'unter Professor and have been supported by the projects CEX2024-001451-M (Unidad de Excelencia ``Mar\'ia de Maeztu'') and PID2023-147112NB-C21, all financed by MICIU/AEI/10.13039/501100011033/ and FEDER, UE.
\end{acknowledgement}

%=========================================================================

\end{document}